\documentclass[aps,pra,twocolumn,superscriptaddress,showpacs,showkeys,amsmath,amssymb]{revtex4}

\usepackage{amsfonts}
\usepackage{amssymb,amsmath}
\usepackage{mathrsfs}
\usepackage{latexsym}
\usepackage{amsmath}
\usepackage[cp1251]{inputenc}
\usepackage{graphicx}
\usepackage{dcolumn}
\usepackage{bm}
\usepackage{color}

\RequirePackage{ifthen}
\RequirePackage[pdfstartview=FitH]{hyperref}
\begin{document}
	
	\title{Efimov-like physics in fraction-dimensional Bose systems with three-body interaction}
	\author{O.~Hryhorchak}
	\author{V.~Pastukhov\footnote{e-mail: volodyapastukhov@gmail.com}}
	\affiliation{Professor Ivan Vakarchuk Department for Theoretical Physics, Ivan Franko National University of Lviv, 12 Drahomanov Street, Lviv, Ukraine}

	\date{\today}

	
	
	\begin{abstract}
	A few-body properties of spinless Bose particles interacting via the contact three-body potential in geometries with fractional dimensions $1<d<2$ are considered. We predict the existence of infinite tower of the Efimov bound states in the four-body sector at three-body resonance. A similar behavior is found in the five-body problem. It is shown that the ratio of high-energy levels in these two sectors is a universal constant. The consequences of emergence of the Efimov physics on the many-body behavior are briefly discussed.
	\end{abstract}
	
	\maketitle
\section{Introduction}
The low-dimensional quantum systems of interacting particles are the excellent platforms for the exploration of various aspects of a few-body correlations and their impact on the macroscopic behavior. Typically described on the microscopic level by the Hamiltonians with the two-body potentials, these systems also serve as an intrinsic playground for testing field theories with the higher-order local self-interactions, which are known to be renormalizable in low dimensions. The appropriate effective field theories are believed to capture the universal low-energy physics of dilute systems on the length scales much larger than a typical ranges of microscopic interaction potentials.

In recent years, the one-dimensional a few- and many-particle systems with the suppressed two-body interaction attracted much attention in the literature \cite{Sowinski_Garcia-March}. The problem of the $SU(3)$ fermions with attractive three-body contact interaction was extensively discussed in articles of Drut et al. \cite{Drut_18,McKenney_19,Czejdo_20}. It was shown that this system features a scale anomaly, which impacts the thermodynamics \cite{McKenney_20} and exact universal relations generalizing the Tan contact theorems for a three-body forces. The quantum anomaly also affects \cite{Maki_19} the high-temperature transport properties of the three-component Fermi gases.

An impact of the three-body interaction on the properties of few bosons in one dimension were studied in Refs.~\cite{Nishida_18,Guijarro_et_al}, and a detailed consideration of purely confinement-induced trimers was performed in \cite{Pricoupenko_18}. The many-body limit is thermodynamically stable only when the interaction between bosons is effectively repulsive. And there are a few ways discussed in the literature for the realization of this repulsion \cite{Valiente_19,Pricoupenko_Petrov_19,Pricoupenko_Petrov_21}. Particularly in Ref.~\cite{Valiente_19}, Valiente proposed a possible one-dimensional scenario for the emergence of the three-body forces among bosons originally interacting through the realistic pairwise potential with infinite scattering length. In the absence of contact term, the residual finite-range effects of the two-body interaction then leads to the effective three-body coupling between particles. A somewhat similar situation can be realized \cite{Pricoupenko_Petrov_19} in the two-channel model near a narrow two-body zero crossing in any spatial dimension.

The short-range character of the three-body repulsion between particles allows the existence of a set of the Tan-like relations for Bose systems \cite{Pastukhov_19}, and simplifies the perturbative calculations for a dilute gas. The one-dimensional bosons with three-body interaction possess a quantum anomaly that influences the many-body properties, namely, shifts \cite{Valiente_Pastukhov} the frequency of the lowest compressional mode of harmonically trapped bosons. The droplet formation and the quantum liquid state at zero temperature in one-dimensional Bose gases with two- and three-body interaction were investigated in \cite{Morera_21}, utilizing low-energy theories and numerical computations. The quantum many-body droplet states of bosons with weak three-body attraction were previously studied in \cite{Sekino_18}.

Important feature of systems with pairwise interactions in the dimensions higher than two is the existence of a non-trivial few-body physics (see \cite{Naidon_Endo_17,Greene_et_al} for recent review), for instance, the universal bound states known as the Efimov effect \cite{Efimov_70}. In one dimension, the infinite tower of universal states emerges \cite{Nishida_Son_10} starting from five particles resonantly interacting through the four-body short-range potential, without the two- and three-body forces involved. In fractional dimensions between one and two, however, the universal physics in a system of bosons with resonant three-body interaction can potentially occur in the four- and five-body sectors, and the present paper addresses the search for these bound states and the study of their impact on the many-body behavior. Although the idea of fractional dimensions is not new for the quantum field theory, where the regularization of the Feynman diagrams typically requires the analytic continuation to non-integer $d$s or for the calculations of the Wilson-Fisher fixed points one utilizes the $\epsilon$-expansion technique, the case of non-relativistic theories below two dimensions is of particular interest. First of all, the three-body terms in $d<2$ are more relevant than a local finite-range corrections to the two-body interactions; secondly, all predictions of these effective field theories can be potentially tested in future experiments with the cold-atoms setup loaded in porous media which support a fractal structure.

\section{Problem statement}
In the following, we discuss properties of bosons with the suppressed contact two-body interactions in a fractional spatial dimensions between $d=1$ and $d=2$. Then, because of the effects of a finite potential range \cite{Valiente_19,Pricoupenko_Petrov_19}, the simplest interaction that survives is the three-body one. Therefore, the three-body (pseudo-) potential has to be taken contact-like $\Phi({\bf r}_1,{\bf r}_2,{\bf r}_3)=g_{3,\Lambda}\delta({\bf r}_{12})\delta({\bf r}_{13})$ with a bare coupling $g_{3,\Lambda}$ that depends on the ultraviolet (UV) cutoff $\Lambda$. The $d$-dimensional $\delta$-functions in $\Phi({\bf r}_1,{\bf r}_2,{\bf r}_3)$ should be understood as some smeared, on the length scale $\Lambda^{-1}$, functions [the simplest choice adopted here for the case of periodic boundary conditions is the following: $\delta({\bf r})=\frac{1}{L^d}\sum_{|{\bf k}|<\Lambda}e^{i{\bf k}{\bf r}}$]. An appropriate (Euclidean) Lagrangian density of the system is specified as follows
\begin{eqnarray}\label{L_bare}
\mathcal{L}=\psi^*\{\partial_{\tau}-\varepsilon\}\psi-\frac{1}{3!} g_{3,\Lambda}(\psi^*)^3\psi^3,
\end{eqnarray}
where $x=(\tau, {\bf r})$, complex field $\psi(x)$ lives in $(d+1)$-dimensional space with the periodic boundary conditions adopted; $\varepsilon=-\frac{\hbar^2\nabla^2}{2m}$ is the one-particle dispersion, and in the thermodynamic limit one should add to (\ref{L_bare}) term with the chemical potential $\mu$.
\subsection{Three-body limit}
\begin{figure}[h!]
	\centerline{\includegraphics
		[width=0.25
		\textwidth,clip,angle=-0]{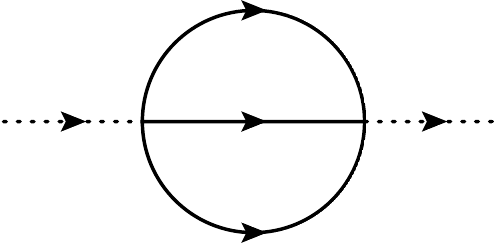}}
	\caption{Diagram determining $\Pi_3(P)$. Solid line denotes the atom propagator.}\label{Pi_3_fig}
\end{figure}

Before proceeding with the discussion of the thermodynamic limit, it is instructive to consider the few-body limits of the theory. Particularly, the three-body one allows to find out the appropriate renormalization of the coupling constant $g_{3,\Lambda}$. For these purposes, let us rewrite Lagrangian (\ref{L_bare}) (generalized to many-body case) in a more convenient nonetheless equivalent way
\begin{eqnarray}\label{L}
\mathcal{L}=\psi^*\{\partial_{\tau}-\varepsilon+\mu\}\psi+\frac{g^{-1}_{3,\Lambda}}{3!} \Psi^*\Psi\nonumber\\
-\frac{1}{3!} \{\Psi^*\psi^3+\textrm{c.c.}\},
\end{eqnarray}
by explicitly introducing an auxiliary complex field $\Psi(x)$ of trimers. With the chemical potential $\mu$ set to zero (zero-density limit), the $\Psi^*\Psi$-correlator can be exactly calculated and reads
\begin{eqnarray}\label{Psi*Psi}
-3!\langle \Psi^*_P\Psi_P\rangle^{-1}=g^{-1}_{3,\Lambda}+\Pi_3(P),
\end{eqnarray}
in the $(d+1)$-momentum space [from now on capital letters denote vectors in the Fourier space, i.e. $P=(\omega_p,{\bf p})$]. The divergent, in the limit $\Lambda \to \infty$, quantity $\Pi_3(P)$ is given by the diagram in Fig.~\ref{Pi_3_fig}
and equals to
\begin{eqnarray}\label{Pi_3}
\Pi_3(P)=\frac{1}{L^{2d}}\sum_{{\bf k}, {\bf q}}\frac{1}{\varepsilon_{k}+\varepsilon_{q}+\varepsilon_{|{\bf k}+{\bf q}+{\bf p}|}-i\omega_p},
\end{eqnarray}
where bold letters denote only spatial part of $(d+1)$-momenta, and both $k=|{\bf k}|$ and $q=|{\bf q}|$ are restricted from the above by $\Lambda$. To regularize this expression, we define bare coupling $g^{-1}_{3,\Lambda}$ in a way that the trimer propagator remains cutoff-independent, i.e. $\frac{d}{d \Lambda}\langle \Psi^*_P\Psi_P\rangle=0$. This procedure is equivalent to the
introduction of the observable coupling $g_3$ by means of formula
\begin{eqnarray}\label{g_3}
g^{-1}_{3}=g^{-1}_{3,\Lambda}+\frac{1}{L^{2d}}\sum_{{\bf k}, {\bf q}}\frac{1}{\varepsilon_{k}+\varepsilon_{q}+\varepsilon_{|{\bf k}+{\bf q}|}}.
\end{eqnarray}
On the dimensional grounds one can introduce the dimensionless (running) three-body coupling $\hat{g}_3\propto g_{3,\Lambda}m\Lambda^{2(d-1)}/\hbar^2$, where the constant prefactor can be chosen in such a way that the renormalization group (RG) equation for $\hat{g}_3$ reads
\begin{eqnarray}
\Lambda\frac{d\hat{g}_3}{d \Lambda}=2(d-1)\hat{g}_3(\hat{g}_3+1).
\end{eqnarray}
The infrared (IR) $\hat{g}_3=0$ and UV $\hat{g}_3=-1$ correspond to the non-interacting limit $g_3=0$ and the unitarity one $g_3=\pm \infty$, respectively. By requiring the $\Psi$-propagator (\ref{Psi*Psi}) to have a real pole at the negative semi-axis after the analytical continuation over the frequency $i\omega_p\to \epsilon_3+i0$, one relates the observable coupling for positive $g_3$s to the three-body vacuum binding energy 
\begin{eqnarray}
g^{-1}_3=-\frac{\Gamma(1-d)}{(2\sqrt{3}\pi)^d}\left(\frac{m}{\hbar^2}\right)^d|\epsilon_3|^{d-1}, \ \ \epsilon_3=-\frac{\hbar^2}{ma^2_3},
\end{eqnarray}
where we parameterized $\epsilon_3$ for positive $g_3$s by the three-body bound state width $a_3$. For convenience, the same parameterization can be used for negative couplings. In both cases $a_3$ coincides with the scattering length. The above-mentioned unitarity points $g_3=\infty$ and $g_3=-\infty$ are associated with the three-body threshold and with the divergency of scattering amplitude at zero energy of the three-body continuum, respectively. Now we can rewrite a formula (\ref{Psi*Psi})
\begin{eqnarray}
-3!\langle \Psi^*_P\Psi_P\rangle^{-1}=g^{-1}_{3}\left[\frac{g_3}{|g_3|}-\frac{(\varepsilon_p/3-i\omega_p)^{d-1}}{|\epsilon_3|^{d-1}}\right],
\end{eqnarray}
after carrying out the integration explicitly. It is readily to show that the r.h.s of the above equation is the inverse three-body $t$-matrix for bosons with the three-body interaction in $d>1$ dimensions. Furthermore, by taking the limit $d\to 1$, we recover the correct expression \cite{Pastukhov_19,Valiente_Pastukhov} even in one dimension.

\subsection{Atom-trimer sector}\label{atom-trimer_subsec}
Lagrangian $\mathcal{L}$ written down through the trimer field $\Psi(x)$, allows for a very simple calculations of the atom-trimer, two-atoms-trimer and the trimer-trimer scattering properties. Below, we mainly focus on a two former cases in the so-called scaling limit, because the correct description of more than the four-body physics requires the introduction of a finite range for the three-body potential.

\begin{figure}[h!]
	\centerline{\includegraphics
		[width=0.5
		\textwidth,clip,angle=-0]{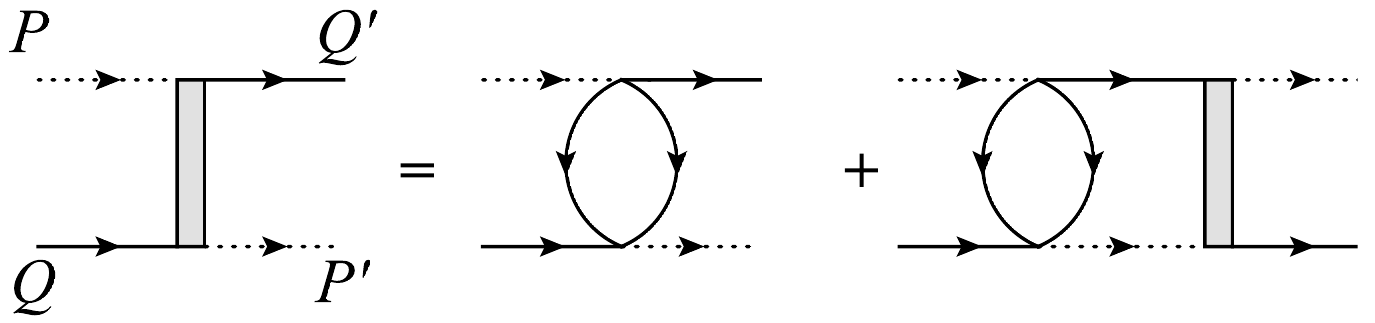}}
	\caption{Integral equation for the atom-trimer vertex $\mathcal{T}_{3-1}(P;Q|Q';P')$ (gray rectangle). Dotted line denotes the trimer propagator.}\label{Tau_3-1_fig}
\end{figure}

\begin{figure}[h!]
	\centerline{\includegraphics
		[width=0.6
		\textwidth,clip,angle=-0]{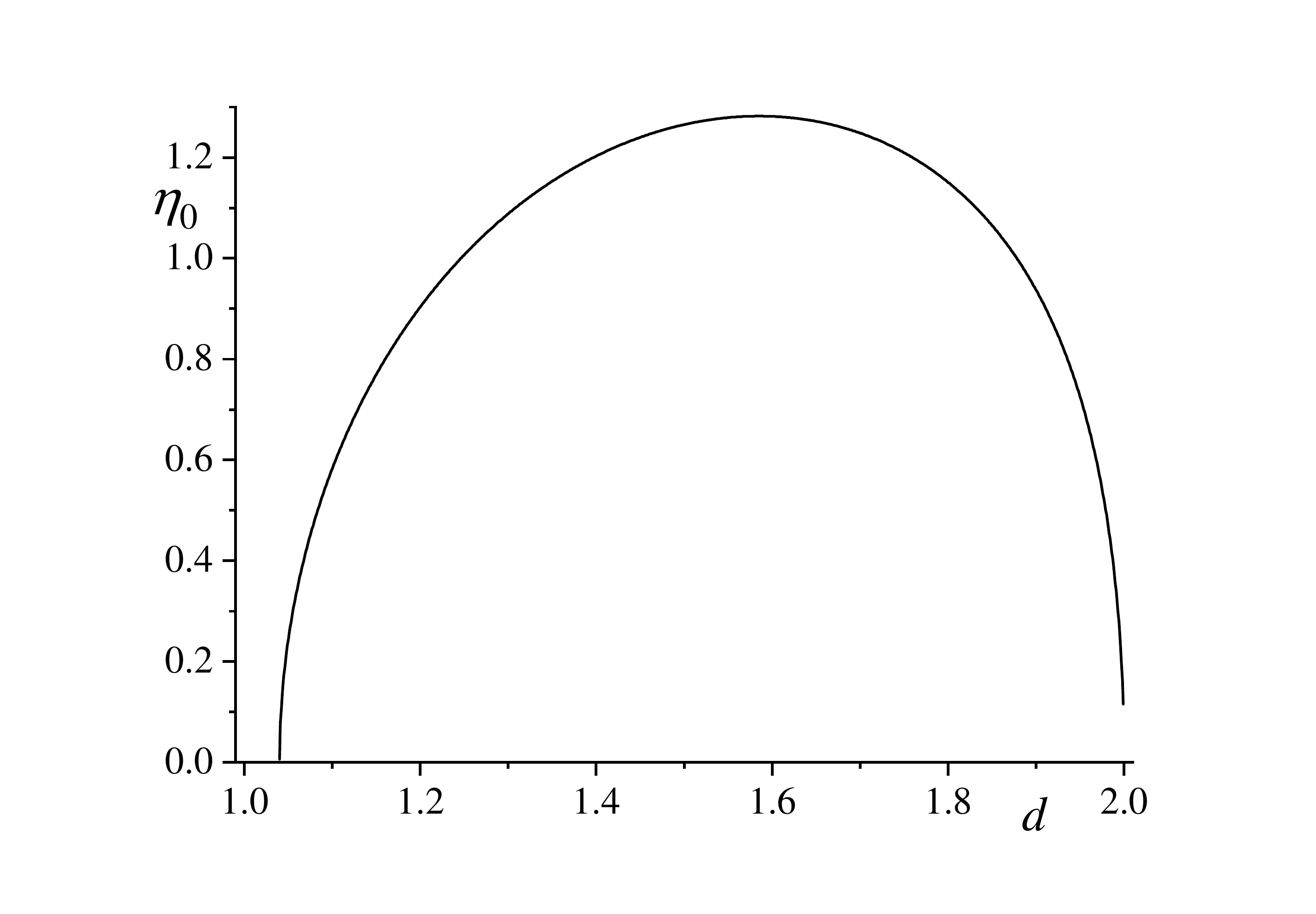}}
	\caption{Numerical solution of algebraic equation (\ref{algebraic_Eq}) for imaginary $\eta=i\eta_0$ in various spatial dimensions. The Efimov-like physics appears only in the window between $d\approx 1.04$ and $d=2$. Recall that for the three-boson system with resonant two-body interaction the Efimov effect emerges only in $2.30<d<3.76$ \cite{Nielsen_01}, with exponent $\eta_0$ of order unity. Exactly in $d=1$, the Efimov effect is allowed \cite{Nishida_Son_10} in the five-body sector for bosons with a four-body forces.}\label{eta_0_fig}
\end{figure}

The diagrams, contributing to the atom-trimer vertex conventionally organize in a linear integral equation presented in Fig.~\ref{Tau_3-1_fig}.
At zero density, the only contribution to the frequency integral comes from the pole of atom propagator $\langle\psi^*_K\psi_K\rangle$, if we encircle a contour in the lower complex half-plane
\begin{eqnarray}\label{int_Eq}
&&\mathcal{T}_{3-1}(P;Q|Q';P')=-\frac{1}{2}\Pi_2(P-Q')\nonumber\\
&&+\frac{1}{L^d}\sum_{{\bf k}}\frac{1}{2}\left\{\Pi_2(P-K)\langle|\Psi_{P+Q-K}|^2\rangle\right.\nonumber\\
&&\left.\times\mathcal{T}_{3-1}(P+Q-K;K|Q';P')\right\}|_{i\omega_k \to \varepsilon_k},
\end{eqnarray}
where the elementary bubble, describing a direct atom-atom scattering process after the carrying out the frequency integration, reads
\begin{eqnarray}\label{Pi_2}
&&\Pi_2(P)=\frac{1}{L^{d}}\sum_{{\bf k}}\frac{1}{\varepsilon_{k}+\varepsilon_{|{\bf k}+{\bf p}|}-i\omega_p}.
\end{eqnarray}
The integral equation (\ref{int_Eq}) simplifies for a vertex with the external lines put on-shell $i\omega_q\to \varepsilon_q$ and $i\omega_{q'}\to \varepsilon_{q'}$ in the center-of-mass frame ${\bf p}=-{\bf q}$ and ${\bf p}'=-{\bf q}'$. In order to preserve the energy conservation, we replace $i\omega_p=i\omega-\varepsilon_q$ (and same for the primed variables), where $i\omega$ should be associated, after analytical continuation, with the (conserved) atom-trimer energy. For the on-shell vertex, $\mathcal{T}_{3-1}({\bf q},{\bf q}')$, we obtain
\begin{eqnarray}\label{int_Eq_simple}
&&\mathcal{T}_{3-1}({\bf q},{\bf q}')=-\frac{1}{2}\Pi_2({\bf q},{\bf q}')\nonumber\\
&&-\frac{1}{L^d}\sum_{{\bf k}}\frac{1}{2}\Pi_2({\bf q},{\bf k})3!t_3(k)\mathcal{T}_{3-1}({\bf k},{\bf q}'),
\end{eqnarray}
with the short-hand notations adopted for a functions
\begin{eqnarray}
&&\Pi_2({\bf q},{\bf k})=\frac{\Gamma\left(1-d/2\right)}{(4\pi)^{d/2}}\left(\frac{m}{\hbar^2}\right)^{d/2}\nonumber\\
&&\times\left(\varepsilon_{|{\bf q}+{\bf k}|}/2+\varepsilon_q+\varepsilon_k-i\omega\right)^{d/2-1},\\
&&t^{-1}_3(k)=g^{-1}_{3}\left[1-\frac{(4\varepsilon_k/3-i\omega)^{d-1}}{|\epsilon_3|^{d-1}}\right].
\end{eqnarray}
It should be noted that Eq.~(\ref{int_Eq_simple}) is somewhat peculiar. Particularly, the problems with its solution arise in the UV region and one cannot just send $\Lambda\to \infty$. A similar situation happens \cite{BHvK_99_1,BHvK_99_2} in higher dimensions $d>2$ for the three-body systems of bosons with a contact pairwise interaction. The simplest way to track back the ambiguity is to put ${\bf q}'=0$ and make use of the iteration procedure. At the first stage one can take $\mathcal{T}_{3-1}({\bf q},{\bf 0})=\mathcal{T}_{3-1}(q)\approx-\frac{1}{2}\Pi_2({\bf q},{\bf 0})$ and neglect the integral in the r.h.s. In order to calculate the second iteration, we substitute the zero-order result under the integral, and then repeat the procedure. At large $q$s, $\Pi({\bf q},{\bf 0})$ is proportional to $1/q^{2-d}$, while the first iteration gains additional large logarithm $\ln (q)/q^{2-d}$. In general, it is easy to show that to the leading order, the $n$-th iteration behaves like $\ln^n (q)/q^{2-d}+\mathcal{O}(\ln^{n-1} (q)/q^{2-d})$ at UV. This suggests that Eq.~(\ref{int_Eq_simple}), in the region where $q\ll \Lambda$ but at the same time $q\gg 1/a_3$ should be much larger than all others the inverse length scales, reduces to the homogeneous one
\begin{eqnarray}\label{int_Eq_hom}
\mathcal{T}_{3-1}(q)=A_d\int_{k<\Lambda}d{\bf k}\frac{k^{2(1-d)}\mathcal{T}_{3-1}(k)}{(k^2+q^2+2{\bf k}{\bf q}/3)^{1-d/2}},
\end{eqnarray}
with a finite and positive definite constant prefactor
$A_d=-\frac{3!\left(3/2\right)^{2(d-1)}}{(4\pi)^{d/2}}\frac{\Gamma(1-d/2)}{\Gamma(1-d)}$
in $1<d<2$. An intrinsic scale invariance in the intermediate region prompts the possible power-law solution $\mathcal{T}_{3-1}(q)\propto q^{d/2-1+\eta}$. Performing the integrations (setting $\Lambda \to \infty$) in Eq.~(\ref{int_Eq_hom}), we obtain an algebraic equation that determines allowed values of the exponent $\eta$
\begin{eqnarray}\label{algebraic_Eq}
\frac{1}{3}=-\frac{\Gamma\left(\frac{1-d/2-\eta}{2}\right)\Gamma\left(\frac{1-d/2+\eta}{2}\right)}{\left(8/9\right)^{d-1}\Gamma(1-d)\Gamma(d/2)}\nonumber\\
\times _{2}F_{1}\left(\frac{1-d/2-\eta}{2},\frac{1-d/2+\eta}{2};d/2;1/9\right),
\end{eqnarray}
where $_{2}F_{1}\left(a,b;c;z\right)$ is the hypergeometric function \cite{Abramowitz}. Note that if $\eta$ is an arbitrary solution to Eq.~(\ref{algebraic_Eq}), then $-\eta$ is also a solution. Therefore, the general solution to the integral equation (\ref{int_Eq_hom}) is a linear combination of these two partial ones with $\pm \eta$
\begin{eqnarray}\label{Tau_sol}
\mathcal{T}_{3-1}(q)\propto q^{d/2-1}\left[\left(\frac{q}{\Lambda_*}\right)^{\eta}+\left(\frac{q}{\Lambda_*}\right)^{-\eta}\right],
\end{eqnarray}
where $\Lambda_*$ is an arbitrary scale depending on the details of the short-range behavior of the three-body potential. Throughout the possible solutions there are a truly imaginary $\eta=i\eta_0$ ones (see Fig.~\ref{eta_0_fig}). In this case $\mathcal{T}_{3-1}(q)$ is an oscillating function of $q$ that supports the discrete scale invariance $q\to qe^{\pi l/\eta_0}$ (with integer $l$). Thus, the appearance of the imaginary roots in Eq.~(\ref{algebraic_Eq}) signals an emergence of the Efimov-like physics \cite{Efimov_70} in the considered four-body system. Exactly in two dimensions the semi-super Efimov effect \cite{Nishida_17} occurs in the four-body sector of our model. 

It should be noted that the subsequent substitution of the solution (\ref{Tau_sol}) in the r.h.s of Eq.~(\ref{int_Eq_hom}) leaves us with the badly-behaved integral at $\Lambda\to \infty$. In order to treat the problem, one has to remember that the effective Lagrangian $\mathcal{L}$ that governs the properties of the system in general is not complete. In fact, one is free to add to (\ref{L}) any (local) interaction terms allowed by the symmetries. The relevant for our problem choice is the vertex describing the direct atom-trimer two-body ($s$-wave) scattering with a (bare) coupling $g_{3-1,\Lambda}$. The latter means the following
\begin{eqnarray}\label{L_intermed}
\mathcal{L}\to \mathcal{L}-g_{3-1,\Lambda}\Psi^*\psi^*\psi\Psi.
\end{eqnarray}
A structurally similar counterterm has to be included in the Lagrangian of spinless fermions with a $p$-wave type contact two-body interaction in the fractional dimensions \cite{Pastukhov_20}, in order to regularize the three-body problem \cite{Sekino_Nishida_21}. This modification should be accounted [one just has to make a replacement $-\frac{1}{2}\Pi_2(P-Q')\to -\frac{1}{2}\Pi_2(P-Q')+g_{3-1,\Lambda}$] in the integral equations (\ref{int_Eq}) and (\ref{int_Eq_simple}) for the atom-trimer scattering amplitude. Then the $\Lambda$-dependence of coupling $g_{3-1,\Lambda}$ can be calculated demanding the modified on-shell vertex satisfies the differential equation $\frac{\partial}{\partial \Lambda}\mathcal{T}_{3-1}({\bf q},{\bf q}')=0$ in the $q,q'\to 0$ limit. This can be performed only numerically. An approximate prescription for finding a solution of this RG equation can be adopted from Refs.~\cite{BHvK_99_1,BHvK_99_2}, where the effective-field-theory approach to the atom-dimer problem with a local pairwise interaction between three-dimensional bosons was developed. Being generalized to our case, the calculation procedure provides a determination of the bare coupling constant $g_{3-1,\Lambda}$ by requiring the 
integral in Eq.~(\ref{int_Eq_hom}) (after the modification of kernel) to be well-defined at UV after the substitution of $\mathcal{T}_{3-1}(q)$ in the form (\ref{Tau_sol}). The compensation of UV divergences appears when $g_{3-1,\Lambda}$ is specified as follows
\begin{eqnarray}
g_{3-1,\Lambda}=\left(\frac{2}{3}\right)^{1-d/2}\frac{\Gamma(1-d/2)}{(8\pi)^{d/2}}\frac{m\hat{g}_{3-1}}{\hbar^2\Lambda^{2-d}},
\end{eqnarray}
where the dimensionless atom-trimer coupling reads
\begin{eqnarray}\label{g_3_1}
\hat{g}_{3-1}=-\frac{\cosh\left[\eta\ln(\Lambda/\Lambda_*)+\delta\right]}{\cosh\left[\eta\ln(\Lambda/\Lambda_*)-\delta\right]},
\end{eqnarray}
with $\delta$ given by $\delta=\textrm{artanh}\frac{\eta}{1-d/2}$. It is instructive \cite{Mohapatra_18}, in a context of emergence of the Efimov effect, to obtain the RG equation for the running atom-trimer coupling
\begin{eqnarray}\label{RG_g_3_1}
\Lambda\frac{d \hat{g}_{3-1}}{d\Lambda}=\frac{2-d}{4}(1+\hat{g}_{3-1})^2-\frac{\eta^2}{2-d}(1-\hat{g}_{3-1})^2.
\end{eqnarray}
For real $\eta$s the beta-function in Eq.~(\ref{RG_g_3_1}) supports two fixed points $\frac{1-d/2\pm\eta}{d/2-1\pm\eta}$, while for the truly imaginary exponents $\eta=i\eta_0$ the fixed points disappear in the complex plane. This is one of three generic scenarios \cite{Kaplan_et_al} for the conformality loss, when UV and IR fixed points of the running atom-trimer coupling annihilate each other, which actually realizes in the presence of the Efimov-like physics. It should be noted that the form of a dimensionless coupling with imaginary $\eta$s supposes the invariance of $\hat{g}_{3-1}$ under the scale transformations $\Lambda\to \Lambda e^{\pi l/\eta_0}$ with integer $l$s. Although we do not discuss here the four-body bound states in details, because this problem requires the modification of the three-body potential at the origin, from the analysis above it should be clear that the scaling limit ($q\ll \Lambda$, $q\gg 1/a_3$) of the atom-trimer bound state wave functions in a momentum space is described by Eq.~(\ref{int_Eq_hom}). And the appropriate energy levels $\epsilon_{3-1}(l)\propto \frac{\hbar^2\Lambda^2_*}{m}e^{-2\pi l/\eta_0}$ display the discrete scale invariance, while the number of these four-body bound states is infinite exactly at the unitarity point $1/a_3=0$. At maximal $\eta_0\approx1.28$ that corresponds to $d\approx 1.59$, the ration of two nearest energy levels is $\epsilon_{3-1}(l)/\epsilon_{3-1}(l+1)=e^{2\pi/\eta_0}\approx (11.6)^2$. For comparison, the scaling factor for three resonantly-interacting bosons in three dimension reads $(22.7)^2$ \cite{Efimov_70}.

\subsection{Two atoms $+$ one trimer}
\begin{figure}[h!]
	\centerline{\includegraphics
		[width=0.45
		\textwidth,clip,angle=-0]{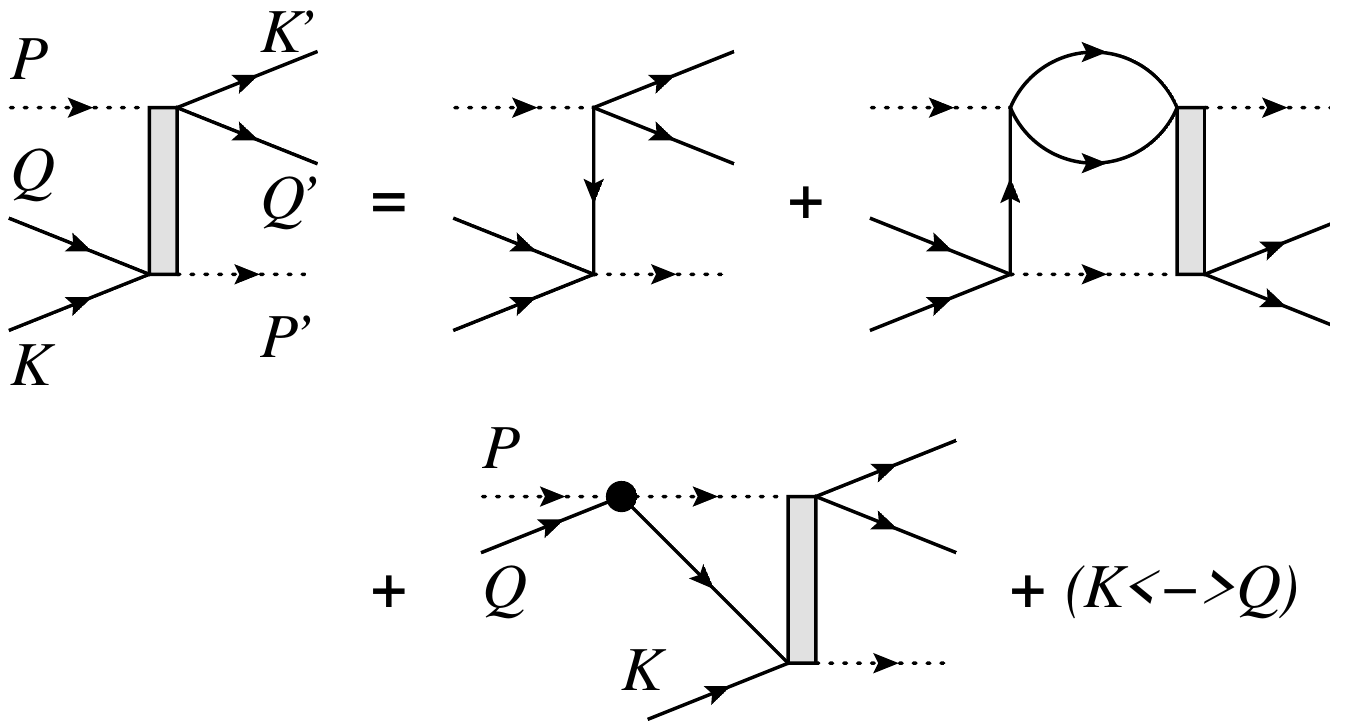}}
	\caption{Equation determining $\mathcal{T}_{3-2}(P;Q,K|K',Q';P')$. The definition of lines is the same as in Figs.~\ref{Pi_3_fig},\ref{Tau_3-1_fig}: solid and dotted lines stand for atom and trimer propagators, respectively. A bold dot stands for sum, $g_{3-1,\Lambda}-\frac{1}{2}\Pi_2(P-Q)$ (see \ref{atom-trimer_subsec}), i.e. for two simplest diagrams for the atom-trimer vertex. }\label{Tau_3-2_fig}
\end{figure}

The five-body problem in our treatment with the trimer fields explicitly involved in the Lagrangian is also tractable. The set of diagrams contributing to the two-atom-trimer vertex $\mathcal{T}_{3-2}(P;Q,K|K',Q';P')$ yields the integral equation presented in Fig.~\ref{Tau_3-2_fig}.
In order to move forward, we integrate over frequencies in the r.h.s., pass to the center-of-mass frame ${\bf p}=-{\bf k}-{\bf q}$ and ${\bf p}'=-{\bf k}'-{\bf q}$, and consider only the on-shell (all $i\omega_{q,k}$s are replaced by $\varepsilon_{q,k}$s) expression (i.e. the two-atoms-trimer scattering amplitude)
\begin{eqnarray}\label{int_Eq_23}
	&&\mathcal{T}_{3-2}({\bf q},{\bf k}|{\bf k}',{\bf q}')=\mathcal{T}^{(0)}_{3-2}({\bf q},{\bf k}|{\bf k}',{\bf q}')\nonumber\\
	&&+\frac{1}{2L^{2d}}\sum_{{\bf p},{\bf s}}\mathcal{T}^{(0)}_{3-2}({\bf q},{\bf k}|{\bf p},{\bf s})3!t_3({\bf p},{\bf s})\mathcal{T}_{3-2}({\bf s},{\bf p}|{\bf k}',{\bf q}')\nonumber\\
	&&+\frac{1}{L^{d}}\sum_{{\bf p}}\left[g_{3-1,\Lambda}-\frac{1}{2}\Pi_2({\bf q},{\bf k},{\bf p})\right]\left[3!t_3({\bf k},{\bf p})\right.\nonumber\\
	&&\left.\times\mathcal{T}_{3-2}({\bf k},{\bf p}|{\bf k}',{\bf q}')+3!t_3({\bf q},{\bf p})\mathcal{T}_{3-2}({\bf q},{\bf p}|{\bf k}',{\bf q}')\right],
\end{eqnarray}
where $\mathcal{T}^{(0)}_{3-2}({\bf q},{\bf k}|{\bf k}',{\bf q}')=\frac{-1}{\varepsilon_{|{\bf k}+{\bf q}+{\bf k}'+{\bf q}'|}+\varepsilon_k+\varepsilon_q+\varepsilon_{k'}+\varepsilon_{q'}-i\omega}$, the energy conservation implies $i\omega_p=i\omega-\varepsilon_{k}-\varepsilon_{q}$, and other notations are a natural generalization of the above formulas
\begin{eqnarray}
&\Pi_2({\bf q},{\bf k},{\bf p})=\frac{\Gamma\left(1-d/2\right)}{(4\pi)^{d/2}}\left(\frac{m}{\hbar^2}\right)^{d/2}\nonumber\\
&\times\left(\varepsilon_{|{\bf q}+{\bf k}+{\bf p}|}/2+\varepsilon_q+\varepsilon_k+\varepsilon_p-i\omega\right)^{d/2-1},\\
&t^{-1}_3({\bf k},{\bf q})=g^{-1}_{3}\left[1-\frac{(\varepsilon_{|{\bf k}+{\bf q}|}/3+\varepsilon_k+\varepsilon_q-i\omega)^{d-1}}{|\epsilon_3|^{d-1}}\right].
\end{eqnarray}
Being interested in the five-body bound states, we can represent the scattering amplitude near its pole $\epsilon_{3-2}$ as $\mathcal{T}_{3-2}({\bf q},{\bf k}|{\bf k}',{\bf q}')=3!t_3({\bf q},{\bf k})Z({\bf q},{\bf k})3!t_3({\bf q}',{\bf k}')Z({\bf q}',{\bf k}')/(i\omega-\epsilon_{3-2})$ [here factors $3!t_3({\bf q},{\bf k})$ are introduced in order to simplify the equation on function $Z({\bf q},{\bf k})$]. Then, the non-homogeneous term in (\ref{int_Eq_23}) can be neglected and we obtain
\begin{eqnarray}\label{int_Eq_Z}
\frac{1}{3!}t^{-1}_3({\bf q},{\bf k})Z({\bf q},{\bf k})=
\frac{1}{2L^{2d}}\sum_{{\bf p},{\bf s}}\mathcal{T}^{(0)}_{3-2}({\bf q},{\bf k}|{\bf p},{\bf s})Z({\bf s},{\bf p})\nonumber\\
+\frac{1}{L^{d}}\sum_{{\bf p}}\left[g_{3-1,\Lambda}-\frac{1}{2}\Pi_2({\bf q},{\bf k},{\bf p})\right]
\left[Z({\bf k},{\bf p})+Z({\bf q},{\bf p})\right],
\end{eqnarray}
where $i\omega$ is replaced everywhere by $\epsilon_{3-2}$. Although we did not success in finding a solution to this equation, an important general physical conclusions, nonetheless, can be drawn. Particularly, by considering the scaling limit of Eq.~(\ref{int_Eq_Z}), it is easy to figure out that, since there is no length scale in the problem, solution $Z({\bf q},{\bf k})$ is expected to be a homogeneous function of its arguments. Furthermore, the limit $q\gg k$ (and vise versa) is fully determined by the atom-trimer problem. Indeed, the large difference between modulus ${\bf k}$ and ${\bf q}$ means that the largest scale (let say $q$) controls the physics of the system. Considering the spacial distribution of two atoms, we immediately conclude that one particle is located much closer (at distances of order $q^{-1}$) to the trimer and consequently forms a universal four-body bound state, while another particle (which is located on scales of order $k^{-1}$ from the trimer) is almost free. A truly five-body effects, therefore, appear only when two arguments of $Z({\bf q},{\bf k})$ are of the same order magnitude. The above suggests the following ansatz $Z({\bf q},{\bf k})=\frac{\mathcal{T}_{3-1}(\sqrt{k^2+q^2})}{(k^2+q^2)^{d-1}}z\left(\frac{2{\bf q}{\bf k}}{k^2+q^2}\right)$, where the latter dimensionless function depends on the angle between ${\bf k}$ and ${\bf q}$, and ratio of their absolute values. In such a way, the UV behavior of the solution of Eq.~(\ref{int_Eq_Z}) in the scaling region ($k,q\ll \Lambda$) is fixed. Plugging the ansatz back into the integral equation, we see that the second and third terms in the r.h.s., because of $g_{3-1, \Lambda}$, are well-behaved at UV, while the first one is not. In order to cure the first (double) integral in (\ref{int_Eq_Z}), we are forced to modify the initial Lagrangian with one more local term that preserves the global $U(1)$ symmetry
\begin{eqnarray}\label{L_final}
\mathcal{L}\to (\ref{L_intermed})-\frac{1}{4}g_{3-2,\Lambda}\Psi^*(\psi^*)^2\psi^2\Psi,
\end{eqnarray}
and describes the direct two-atom-trimer interaction with a bare coupling $g_{3-2,\Lambda}$. The dimensional analysis fixes the overall factor $g_{3-2,\Lambda}\propto \frac{m\hat{g}_{3-2}}{\hbar^2\Lambda^{2}}$, while the explicit dependence of a running coupling $\hat{g}_{3-2}$ can be obtained by making the low-momentum limit of the two-atom-trimer scattering amplitude (\ref{int_Eq_23}) insensitive to the UV cutoff. This is, of course, a very complicated numerical problem and here, instead, we calculate $\hat{g}_{3-2}$ by utilizing the previously described approximate procedure demanding the contribution of upper integration limit to be vanishing. The details of calculations are presented in Appendix, though the final result is simple
\begin{eqnarray}\label{g_3_2}
\hat{g}_{3-2}=-\frac{\cosh\left[\eta\ln(\Lambda/\Lambda_*)+\delta_u\right]}{\cosh\left[\eta\ln(\Lambda/\Lambda_*)-\delta_d\right]},
\end{eqnarray}
where universal phases $\delta_u$ and $\delta_d$ functionally depend (see Appendix) on the solution of Eq.~(\ref{int_Eq_Z}) in the scaling region. Likewise $\hat{g}_{3-1}$, the RG beta-function for coupling $\hat{g}_{3-2}$ at real $\eta$s possesses two zeros corresponding to UV and IR fixed points, respectively. For the Efimov region, the phases are imaginary $\delta_u=i\delta_{u0}$ and $\delta_d=i\delta_{d0}$ and $\hat{g}_{3-2}$, as the function of the UV cutoff diverges at discrete infinite set of points signaling an emergence of the five-body bound states. By comparing Eq.~(\ref{g_3_2}) and Eq.~(\ref{g_3_1}), we can obtain an universal ratio between the two-atom-trimer and the atom-trimer bound-state energies
\begin{eqnarray}
\frac{\epsilon_{3-2}(l)}{\epsilon_{3-1}(l)}=e^{2(\delta_{d0}-\delta_{0})/\eta_0},
\end{eqnarray}
valid for the large $l$ close to the unitarity point. The observed physical picture is a somewhat similar to a system of bosons with the resonant two-body interaction in three dimensions, where the energy levels of the $l$-th Efimov's trimer and $l$-th tetramer are related to each other by the universal scaling factor \cite{Platter_04,Hammer_07,vonStecher_09}, the so-called Tjon line \cite{Tjon_75}. Actually, this resemblance is not totally unexpected, because kinematically our problem is very similar to the bosonic system with pairwise inter-particle interaction in the dimensions between $d=2$ and $d=4$. An open question is the number of different branches of the five-body bound states in the considered model. The answer can be given by a more sophisticated consideration of problem with the realistic three-body potential, since there are some arguments \cite{Schmidt_10} that the effective field theory captures only the deepest states.

\section{Contacts and Tan-like energy relation}
Results from the previous section should have a profound effect on the many-body behavior of the system. In particular, they fix, through the set of the density-dependent constants, the universal high-momentum tail of the particle distribution. Taking into account the previous findings \cite{Pastukhov_19} in the one-dimensional case, one can naively generalize the leading-order asymptotics, $N_p=\mathcal{C}_3/p^{4-d}$, where $\mathcal{C}_3$ is the three-body contact parameter. In higher dimensions $d>1.04$, however, we should expect that due to emergence of the Efimov physics, the next-to-leading-order log-periodic terms appear.

Let us obtain a formally exact expression for the ground-state energy of a system. These calculations can be done in the two-fold way, each refer to the atom-trimer Lagrangian density. The first one exploits \cite{Pastukhov_2d} the Hellmann-Feynman theorem. Another method utilizes the exact identities following from the equations of motion for various correlation functions. By recognizing $g_{3}$ as a measurable three-body coupling parameter, the internal energy of bosons can be computed as a sum of derivatives of the thermodynamic potential with respect to the mass of particles and coupling
$E=\left(-m\frac{\partial }{\partial m}+g_3\frac{\partial }{\partial g_3}\right)_{\mu}\Omega$,
or rewriting in terms of the (thermal) average $\langle \ldots \rangle$
\begin{eqnarray}
E/L^d=\left(m\frac{\partial }{\partial m}-g_3\frac{\partial }{\partial g_3}\right)_{\mu}\langle\mathcal{L}\rangle,
\end{eqnarray}
where $\mathcal{L}$ should be identified with Eq.~(\ref{L_final}). While calculating the kinetic energy term, one has to keep in mind the explicit dependence on mass not only in $\varepsilon$ but also in $g_{3,\Lambda}$, $g_{3-1,\Lambda}$ and $g_{3-2,\Lambda}$ as well
\begin{eqnarray}\label{Kin_en}
m\frac{\partial }{\partial m}\langle\mathcal{L}\rangle=\langle\psi^*\varepsilon\psi\rangle+\frac{1}{3!}\left(g^{-1}_{3,\Lambda}-g^{-1}_{3}\right)\langle\Psi^*\Psi\rangle\nonumber\\
-g_{3-1,\Lambda}\langle\Psi^*\psi^*\psi\Psi\rangle-\frac{1}{4}g_{3-2,\Lambda}\langle\Psi^*(\psi^*)^2\psi^2\Psi\rangle.
\end{eqnarray}
A factor near the second term in r.h.s. of the equation above is the divergent sum from Eq.~(\ref{g_3}). The same divergence, but with an opposite sign appears in the course of calculations of the first term, $\langle\psi^*\varepsilon\psi\rangle$, therefore, the energy $E$ remains finite even when the UV cutoff is sent to infinity. This fact immediately identifies the three-body contact parameter for bosons with triple interactions in generic dimension $d<2$
\begin{eqnarray}\label{C_3}
\mathcal{C}_3=\frac{1}{3}\left(\frac{4}{3}\right)^{1-d/2}\frac{\Gamma(1-d/2)}{(4\pi)^{d/2}}\left(\frac{m}{\hbar^2}\right)^{2}\langle\Psi^*\Psi\rangle.
\end{eqnarray}
Being rewritten in terms of single-atom fields on the classical level $\Psi\to \frac{g_{3,\Lambda}}{1-3!g_{3,\Lambda}g_{3-1,\Lambda}\psi^*\psi-\frac{3!}{4}g_{3,\Lambda}g_{3-2,\Lambda}(\psi^*)^2\psi^2}\psi^3$, the three-body contact, in general, contains infinite number of terms. In 1D we can neglect the couplings $g_{3-1,\Lambda}$ and $g_{3-2,\Lambda}$, and obtain a simple result  $\mathcal{C}_3=\frac{1}{3\sqrt{3}}\left(\frac{mg_{3,\Lambda}}{\hbar^2}\right)^{2}\langle(\psi^*)^3\psi^3\rangle$ \cite{Pastukhov_19}. For higher dimensions $d>1.04$, two more contact parameters, associated with terms $g_{3-1,\Lambda}\langle\Psi^*\psi^*\psi\Psi\rangle$ and $\frac{1}{4}g_{3-2,\Lambda}\langle\Psi^*(\psi^*)^2\psi^2\Psi\rangle$ in the internal energy, arise. In order to find out the precise relations of these terms in $E$ to $N_p$, one has to calculate the next-to-leading-order terms in the particle distribution at large momenta. The latter quantity can be calculated utilizing the operator product expansion \cite{Braaten_08} for the reduced single-particle density matrix (equal-time atom propagator) at small distances.

The calculations of the average potential energy density end up with a finite expression
\begin{eqnarray}\label{Pot_en}
-g_3\frac{\partial }{\partial g_3}\langle\mathcal{L}\rangle=\frac{g^{-1}_{3}}{3!}\langle\Psi^*\Psi\rangle.
\end{eqnarray}

The above way for the derivation of Tan's identity is consistent with the scheme that makes use of the equation of motion for field $\Psi$. Indeed, from the atom-trimer Lagrangian (\ref{L_final}) one readily obtains the energy density as follows
\begin{eqnarray}
\langle\psi^*\varepsilon\psi\rangle-\frac{g^{-1}_{3,\Lambda}}{3!}\langle\Psi^*\Psi\rangle+\frac{1}{3!}\{\langle\Psi^*\psi^3\rangle+\textrm{c.c.}\}\nonumber\\
+g_{3-1,\Lambda}\langle\Psi^*\psi^*\psi\Psi\rangle+\frac{1}{4}g_{3-2,\Lambda}\langle\Psi^*(\psi^*)^2\psi^2\Psi\rangle,
\end{eqnarray}
and if we now express the averages $\langle\Psi^*\psi^3\rangle$ (and the complex-conjugated one) via an exact equations of type
\begin{eqnarray}
-\langle \Psi^*\frac{\partial\mathcal{L}}{\partial\Psi'^*}\rangle=\delta(x-x')=-\frac{g^{-1}_{3,\Lambda}}{3!}\langle\Psi^*\Psi'\rangle\nonumber\\
+\frac{1}{3!}\langle\Psi^*\psi'^3\rangle+g_{3-1,\Lambda}\langle\Psi^*\psi'^*\psi'\Psi'\rangle\nonumber\\
+\frac{1}{4}g_{3-2,\Lambda}\langle\Psi^*(\psi'^*)^2(\psi')^2\Psi'\rangle,
\end{eqnarray}
(here prime near fields denotes their dependence on $x'$) we arrive with the previously announced result $(\ref{Kin_en})+(\ref{Pot_en})$ (note that the normal-ordering prescription kills out the $\delta$-term).

\section{Summary}
In conclusion, we have considered a system of bosons with the contact three-body interaction in fractional dimensions between $1<d<2$. After the reformulation of the effective field theory by introducing the trimer fields explicitly, we have renormalized the coupling constant by relating it to the three-body binding energy and UV cutoff. Close to the three-body resonance in $1.04<d<2$, we predicted the existence of the infinite tower of the Efimov-like bound states in the four-body sector. It is shown that the proper renormalization of this problem requires introduction of additional scale associated with the direct (contact) atom-trimer interaction term in the Lagrangian. The emergence of the Efimov-like physics was independently confirmed by the renormalization group analysis of the running atom-trimer coupling. The discrete scale invariance manifests itself by the conformality loss. A very similar patterns were found in the five-body problem. Particularly, it is demonstrated that for the appropriate treatment of the two-atom-trimer sector, one needs to include the higher-order interaction term to the Lagrangian. Importantly, that the latter modification of the initial model does not involve any additional parameter related to the five-body physics. This conjectures that the energy ratio of four- and five-body 
bound states is universal. Finally, we obtained Tan's energy relation for the bosons with the three-body interaction and identified contact parameter governing the leading-order tail of a particle distribution at large momenta.

\begin{center}
	{\bf Acknowledgements}
\end{center}
V.P. thanks Dr. V. Vasyuta for the invaluable encouragement
at the last stage of this work. We are grateful to Dr. I. Pastukhova for useful remarks regarding the manuscript, and Prof. Y. Nishida for pointing out Ref.~\cite{Nishida_17} to our attention.

\section{Appendix}
In this section we review some details of the two-atom-trimer coupling calculations. The full point is to cancel out, by fitting $g_{3-2,\Lambda}$, the contribution of upper integration limit in the double integral of Eq.~(\ref{int_Eq_Z})
\begin{eqnarray*}
\int_{p,s=\Lambda}d{\bf p}d{\bf s}\left[\mathcal{T}^{(0)}_{3-2}({\bf 0},{\bf 0}|{\bf p},{\bf s})+g_{3-2,\Lambda}\right]Z({\bf s},{\bf p})=0.
\end{eqnarray*}
Then, using properties of $Z({\bf s},{\bf p})$ and performing change of variables $s=\rho \cos\varphi$ and $p=\rho \sin\varphi$, where $\rho=\sqrt{s^2+p^2}$ and $\varphi \in [0, \pi/2]$, we have (up to overall factor in the l.h.s.)
\begin{eqnarray*}
	\int^{\pi}_{0}d\theta \sin^{d-2}\theta	\int^{\pi/4}_{0}d\varphi \sin^{d-1}(2\varphi)z(\cos\theta\sin(2\varphi))\\
	\times\int^{\Lambda/\cos\varphi}d\rho \rho^{d/2}\left[\left(\frac{q}{\Lambda_*}\right)^{\eta}+\left(\frac{q}{\Lambda_*}\right)^{-\eta}\right]\\
	\times\left[\frac{\hbar^2g_{3-2,\Lambda}}{m}-\frac{1/\rho^2}{1+\frac{1}{2}\cos\theta\sin(2\varphi)}\right]
	=0,
\end{eqnarray*}
(note that integral over $\varphi$ in the interval $[\pi/4,\pi/2]$ gives identical contribution). The solution
 \begin{eqnarray*}
 g_{3-2,\Lambda}=\sqrt{\frac{C_{+}C_{-}}{B_{+}B_{-}}}\frac{m\hat{g}_{3-2}}{\hbar^2\Lambda^{2}},
 \end{eqnarray*}
as well as the universal phases
 \begin{eqnarray*}
	\delta_u=\frac{1}{2}\ln\frac{C_-}{C_{+}}, \ \ \delta_d=\frac{1}{2}\ln\frac{B_{+}}{B_{-}},
\end{eqnarray*}
can be conventionally written in terms of four constants 
 \begin{eqnarray*}
	B_{\pm}=\int^{\pi}_{0}d\theta \sin^{d-2}\theta	\int^{\pi/2}_{0}d\varphi \sin^{d-1}\varphi\\
	\times z(\cos\theta\sin\varphi)
    \frac{\left(\cos\frac{\varphi}{2}\right)^{-1-d/2\pm \eta}}{1+d/2\mp \eta},
\end{eqnarray*}
\begin{eqnarray*}
	C_{\pm}=\int^{\pi}_{0}d\theta \sin^{d-2}\theta	\int^{\pi/2}_{0}d\varphi \sin^{d-1}\varphi\\
	\times	\frac{z(\cos\theta\sin\varphi)}{1+\frac{1}{2}\cos\theta\sin\varphi}
    \frac{\left(\cos\frac{\varphi}{2}\right)^{1-d/2\pm \eta}}{1-d/2\pm \eta}.
\end{eqnarray*}
that functionally depend on the peculiarities of the five-body physics in the scaling region.


\begin{thebibliography}{99}
	
\bibitem{Sowinski_Garcia-March} T. Sowi{\'{n}}ski and M. {\'{A}}ngel Garc{\'{\i}}a-March, \href{https://doi.org/10.1088/1361-6633/ab3a80}{Rep. Prog. Phys. {\bf 82}, 104401 (2019).}


\bibitem{Drut_18} J. E. Drut, J. R. McKenney, W. S. Daza, C. L. Lin, and C. R. Ord\'o\~nez, \href{https://doi.org/10.1103/PhysRevLett.120.243002}{Phys. Rev. Lett. {\bf 120}, 243002 (2018).}

\bibitem{McKenney_19} J. R. McKenney and J. E. Drut, \href{https://doi.org/10.1103/PhysRevA.99.013615}{Phys. Rev. A {\bf 99}, 013615 (2019).}

\bibitem{Czejdo_20} A. J. Czejdo, J. E. Drut, Y. Hou, J. R. McKenney, and K. J. Morrell, \href{https://doi.org/10.1103/PhysRevA.101.063630}{Phys. Rev. A {\bf 101}, 063630 (2020).}

\bibitem{McKenney_20} J. R. McKenney, A. Jose, and J. E. Drut, \href{https://doi.org/10.1103/PhysRevA.102.023313}{Phys. Rev. A {\bf 102}, 023313 (2020).}

\bibitem{Maki_19} J. Maki and C. R. Ord\'o\~nez, \href{https://doi.org/10.1103/PhysRevA.100.063604}{Phys. Rev. A 100, 063604 (2019).}


\bibitem{Nishida_18} Y. Nishida, \href{https://doi.org/10.1103/PhysRevA.97.061603}{Phys. Rev. A {\bf 97}, 061603(R) (2018).}

\bibitem{Guijarro_et_al} G. Guijarro, A. Pricoupenko, G. E. Astrakharchik, J. Boronat, and D. S. Petrov, \href{https://doi.org/10.1103/PhysRevA.97.061605}{Phys.~Rev.~A {\bf 97}, 061605(R) (2018).}

\bibitem{Pricoupenko_18} L. Pricoupenko, \href{https://doi.org/10.1103/PhysRevA.97.061604}{Phys. Rev. A {\bf 97}, 061604(R) (2018)}.




\bibitem{Valiente_19} M. Valiente, \href{https://doi.org/10.1103/PhysRevA.100.013614}{Phys. Rev. A {\bf 100}, 013614 (2019).}

\bibitem{Pricoupenko_Petrov_19} A. Pricoupenko and D. S. Petrov, \href{https://doi.org/10.1103/PhysRevA.100.042707}{
Phys. Rev. A {\bf 100}, 042707 (2019).}

\bibitem{Pricoupenko_Petrov_21} A. Pricoupenko and D. S. Petrov, \href{https://doi.org/10.1103/PhysRevA.103.033326}{
	Phys. Rev. A {\bf 103}, 033326 (2021).}


\bibitem{Pastukhov_19} V. Pastukhov, \href{https://doi.org/10.1016/j.physleta.2018.12.006}{Phys.
	Lett. A 383, 894 (2019).}

\bibitem{Valiente_Pastukhov} M. Valiente and V. Pastukhov, \href{https://doi.org/10.1103/PhysRevA.99.053607}{Phys. Rev. A {\bf 99}, 053607 (2019).}

\bibitem{Morera_21} I. Morera, B. Julia-Diaz, M. Valiente, \href{https://arxiv.org/abs/2103.16499}{arXiv:2103.16499.}

\bibitem{Sekino_18} Y. Sekino and Y. Nishida, \href{https://doi.org/10.1103/PhysRevA.97.011602}{Phys. Rev. A {\bf 97}, 011602(R) (2018).}


\bibitem{Naidon_Endo_17} P. Naidon and S. Endo, \href{https://doi.org/10.1088/1361-6633/aa50e8}{Rep. Prog. Phys. {\bf 80}, 056001 (2017).}

\bibitem{Greene_et_al} C. H. Greene, P. Giannakeas, and J. P\'erez-R\'{\i}os, \href{https://doi.org/10.1103/RevModPhys.89.035006}{Rev. Mod. Phys. {\bf 89}, 035006 (2017).}

\bibitem{Efimov_70} V. Efimov, \href{https://doi.org/10.1016/0370-2693(70)90349-7}{Phys. Lett. B {\bf 33}, 563 (1970).}

\bibitem{Nishida_Son_10} Y. Nishida and D. T. Son, \href{https://doi.org/10.1103/PhysRevA.82.043606}{Phys.
	Rev. A {\bf 82}, 043606 (2010).}


\bibitem{BHvK_99_1}  P. F. Bedaque, H.-W. Hammer, and U. van Kolck, \href{https://doi.org/10.1103/PhysRevLett.82.463}{Phys. Rev. Lett. {\bf 82}, 463 (1999).}

\bibitem{BHvK_99_2} P. F. Bedaque, H.-W. Hammer, and U. van Kolck, \href{https://doi.org/10.1016/S0375-9474(98)00650-2}{Nucl. Phys. A {\bf 646}, 444 (1999).}

\bibitem{Abramowitz} M. Abramowitz and I. Stegun, {\it Handbook of Mathematical
	Functions with Formulas, Graphs, and Mathematical Tables} (United States Department of Commerce, National
Bureau of Standards 1964).

\bibitem{Nielsen_01} E. Nielsen, D. V. Fedorov, A. S. Jensen, and E. Garrido, \href{https://doi.org/10.1016/S0370-1573(00)00107-1}{Phys. Rep. {\bf 347}, 373 (2001).}

\bibitem{Nishida_17} Y. Nishida, \href{https://doi.org/10.1103/PhysRevLett.118.230601}{Phys. Rev. Lett. {\bf 118}, 230601 (2017).}


\bibitem{Pastukhov_20} V. Pastukhov, \href{https://doi.org/10.1103/PhysRevA.102.013307}{Phys. Rev. A {\bf 102}, 013307 (2020).}

\bibitem{Sekino_Nishida_21} Y. Sekino and Y. Nishida, \href{https://doi.org/10.1103/PhysRevA.103.043307}{Phys. Rev. A {\bf 103}, 043307 (2021).}



\bibitem{Mohapatra_18} A. Mohapatra and E. Braaten, \href{https://doi.org/10.1103/PhysRevA.98.013633}{Phys. Rev. A {\bf 98}, 013633 (2018).}

\bibitem{Kaplan_et_al} D. B. Kaplan, J.-W. Lee, D. T. Son, and M. A. Stephanov, \href{https://doi.org/10.1103/PhysRevD.80.125005}{Phys. Rev. D {\bf 80}, 125005 (2009).} 




\bibitem{Platter_04} L. Platter, H.-W. Hammer, and Ulf-G. Mei\ss{}ner, \href{https://doi.org/10.1103/PhysRevA.70.052101}{Phys. Rev. A {\bf 70}, 052101 (2004).}

\bibitem{Hammer_07} H.-W. Hammer, and L. Platter, \href{https://doi.org/10.1140/epja/i2006-10301-8}{Eur. Phys. J. A {\bf 32}, 113
(2007).}

\bibitem{vonStecher_09} J. von Stecher, J. P. D’Incao, and C. H. Greene, \href{https://doi.org/10.1038/nphys1253}{Nature Phys. {\bf 5}, 417 (2009).}

\bibitem{Tjon_75} J.A.Tjon, \href{https://doi.org/10.1016/0370-2693(75)90378-0}{Phys. Lett. B {\bf 56}, 217 (1975).}

\bibitem{Schmidt_10} R. Schmidt and S. Moroz, \href{https://doi.org/10.1103/PhysRevA.81.052709}{Phys. Rev. A {\bf 81}, 052709 (2010).}




\bibitem{Pastukhov_2d} V. Pastukhov, \href{https://doi.org/10.1007/s10909-018-2082-1}{J. Low Temp. Phys. {\bf 194}, 197 (2019).}

\bibitem{Braaten_08} E.~Braaten and L.~Platter, \href{https://doi.org/10.1103/PhysRevLett.100.205301}{Phys.~Rev.~Lett. {\bf 100}, 205301 (2008).}



\end{thebibliography}
\end{document}